\documentclass[pra,letterpaper,english,reprint,nofootinbib,aps,superscriptaddress,showpacs,showkeys]{revtex4-1}

\usepackage{babel,calc,amsmath,amsthm,amssymb,graphicx,subfigure,xcolor,longtable}

\usepackage{braket}
\usepackage{dsfont}
\usepackage[T1]{fontenc}
\usepackage[unicode=true]{hyperref}
\hypersetup{
     colorlinks=true,       		
     linkcolor=blue,          	
     citecolor=red,            
     urlcolor=magenta,           	
}
\newtheorem*{theorem*}{Theorem}

\newtheorem*{corollary*}{Corollary}

\newtheorem*{lemma*}{Lemma}

\newtheorem*{proposition*}{Proposition}

\newtheorem*{conjecture*}{Conjecture}
\theoremstyle{definition}

\newtheorem*{definition*}{Definition}
\theoremstyle{remark}

\newtheorem*{remark*}{Remark}

\begin{document}
   \renewcommand{\figurename}{FIG.}
	\renewcommand{\figureautorefname}{FIG.}
   \title{Entanglement measurement based on convex hull properties}

   \author{Hao-Nan Qiang}
   \affiliation{Theoretical Physics Division, Chern Institute of Mathematics, Nankai University, Tianjin 300071, People's Republic of China}
   
   \author{Jing-Ling~Chen}
   \email{chenjl@nankai.edu.cn}
   \affiliation{Theoretical Physics Division, Chern Institute of Mathematics, Nankai University, Tianjin 300071,
      People's Republic of China}

   \date{\today}
   \begin{abstract}
      Quantum entanglement is a unique correlation phenomenon in quantum mechanics, and the measurement of quantum entanglement plays an important role in quantum computing and quantum communication. Many mainstream entanglement criteria and measurement methods currently known have shortcomings in certain aspects, such as not being sufficient or necessary conditions for entanglement, or only being effective in simple cases such as 2-qubits or pure states. In this work, we will propose a scheme for measuring quantum entanglement, which starts with treating the set of quantum separable states as a convex hull of quantum separable pure states, and analyzes the properties of the convex hull to obtain a new form of entanglement measurement. Although a large amount of data is required in the measurement process, this method is not only applicable to 2-qubit quantum states, but also a entanglement measurement method that can be applied to any dimension and any fragment. We will provide several examples to compare their results with other entanglement metrics and entanglement determination methods to verify their feasibility.

   \end{abstract}
   \maketitle

   \section{Introduction}
   The concept of quantum entanglement was first proposed by Einstein, Podolski, and Rosen (EPR) in 1935. They introduced the EPR paradox in their famous EPR paper\cite{EPR1935}, questioning whether quantum mechanics can fully describe physical reality. The EPR paradox holds that the existence of quantum entangled states violates the assumption of local reality in classical physics, which states that the properties of a system should be determined within a local region rather than through remote immediate effects. In their conception, quantum entanglement seems to imply "superluminal" information transmission, which contradicts Einstein's expectations of relativity and is therefore considered a major flaw in quantum mechanics.
   
   In 1964, John Bell proposed Bell's inequality and designed a series of experiments to verify the differences between quantum mechanics predictions and classical physics theories\cite{bell1964}. The experimental results of Bell's inequality not only support the predictions of quantum mechanics, but also confirm the authenticity of quantum entanglement. Subsequent experiments, such as the Aspert experiment \cite{aspect1982experimental}, proved the existence of quantum entanglement and further solidified the fundamental role of quantum mechanics in describing the microscopic world.
   
   The measurement of quantum entanglement is an important research field in quantum information science. In the research process of quantum entanglement measurement, entanglement entropy\cite{vedral1997quantifying}, Concurrence\cite{wootters2001entanglement}, and PPT (Positive Partial Transpose) methods\cite{peres1996separability} have played a key role. But their applicability may not be limited to all entanglement cases, such as entanglement entropy only applicable to pure states, the Concur method only applicable to 2qubits, and the PPT method cannot prove the sufficient and necessary conditions for entanglement in all mixed state cases\cite{horodecki1996necessary}.

   There are also some other entanglement measurement methods, such as logarithmic negativity\cite{perlmutter2015positivity}, Wigner function\cite{banerji2014entanglement}, symmetric-measurement-based positive maps\cite{li2024quantum} and quantum variational optimization\cite{diez2021quantum}. Recently, with the development of machine learning, corresponding applications have also been put into the field of quantum entanglement measurement\cite{lin2023quantifying}\cite{greenwood2023machine}.
   
In the second part, we will link entanglement with convex hull and provide the entanglement measurement method at this time. In the third part, we will provide the specific form of the entanglement measure and provide some numerical and analytical calculation methods. In the fourth part, we will provide an example in the case of 2-qubits and compare it with the results of the necessary and sufficient PPT method in the case of 2-qubits\cite{horodecki1996necessary}.

\section{Connect entanglement with convex hull}

We can write the density matrix $\rho$ of a $d\times d$ quantum state as the dot product of a vector $\overrightarrow{R} $ and a series of $d\times d$ matrices $\overrightarrow{\Theta }$\cite{chen2002degree}:

\begin{align}
   \rho=\frac{1}{d}(I+\overrightarrow{R}\cdot \overrightarrow{\Theta })\\
   Tr(\Theta_i)=0\\
   \frac{1}{d}Tr(\Theta_i \cdot \Theta_j)=\delta_{ij}
\end{align}

According to the definition of separable states, any separable states $\rho^{S}$ can be written as a series of convex groups and forms of separable pure states $| \psi^{S} \rangle \langle \psi^{S} |$:

\begin{equation}
   \rho^{S}=\sum_{i}p_i| \psi_i^{S} \rangle \langle \psi_i^{S} |
\end{equation}

Therefore, the vector $\overrightarrow{R} $ representing it can be formed by the convex combination of the vectors $\overrightarrow{r} $  corresponding to all pure separable states. In other words, any vector of a quantum separable state is within the convex hull of the set of quantum separable pure state vectors\cite{bertsekas2003convex}.

This convex hull has a series of properties. Firstly, we can conclude that when all components of the vector are 0, the quantum state becomes a unit matrix $\frac{1}{d}I$, and the unit matrix must be a separable state. So we come to the first conclusion that the zero vector is within the convex hull.

Secondly, if the vector $\overrightarrow{r}$ corresponding to a quantum state is within the convex hull, its vector is shortened by mixing the quantum state with the identity matrix, and the vector $p\overrightarrow{r}(p\in [0,1])$ remains within the convex hull, because of the mixture of two separable states is still separable.

The third point is that any quantum state will become separable and its vector will be in a convex hull after sufficient mixing of the identity matrix, which means that its vector length will be shortened to a certain extent. We will use the 2-qubit case to illustrate this point, and similar methods can be used to prove the feasibility of this conclusion in other cases.

In the simple 2-qubit case, we can choose the corresponding matrix $\Theta$ as $\sigma_i \otimes \sigma_j$, the tensor product between Pauli matrices\cite{pauli1988quantenmechanik}:

\begin{equation}
   \sigma_1=I,\sigma_2=\sigma_x,\sigma_3=\sigma_y,\sigma_4=\sigma_z
\end{equation}

\begin{equation}
   \sigma_x=\begin{pmatrix}
       0  & 1  \\
       1  & 0  
   \end{pmatrix},\sigma_y=\begin{pmatrix}
       0  & -i  \\
       i  & 0  
   \end{pmatrix},\sigma_z=\begin{pmatrix}
      1  & 0  \\
      0  & -1  
  \end{pmatrix}
\end{equation}

For any 2-qubit quantum state, we can write the corresponding vector of its density matrix as:

\begin{equation}
   R=\begin{pmatrix}
        & R_{12} & R_{13} & R_{14} \\
       R_{21} & R_{22} & R_{23} & R_{24} \\
       R_{31} & R_{32} & R_{33} & R_{34} \\
       R_{41} & R_{42} & R_{43} & R_{44}
   \end{pmatrix}
\end{equation}

\begin{equation}
   R_{ij}=Tr(\rho\cdot\sigma_i \otimes \sigma_j)
\end{equation}

Since the vector component corresponding to $R_{11}$ is always the coefficient 1 before $\frac{1}{d}I$, we will always omit it from the table.

For a separable 2-qubit quantum pure state, its corresponding vector can be written as\cite{chen2002degree}:

\begin{equation}
   r=\begin{pmatrix}
        & u_x & u_y & u_z \\
       v_x & v_x u_x & v_x u_y & v_x u_z \\
       v_y & v_y u_x & v_y u_y & v_y u_z \\
       v_z & v_z u_x & v_z u_y & v_z u_z
   \end{pmatrix}
\end{equation}

$\overrightarrow{u}$ and $\overrightarrow{v} $ respectively represent the Bloch sphericity loss of each qubit pure states\cite{bloch1946nuclear}, from which we derive the vector form of any 2-qubit quantum separable pure state. 

Taking the element at position $R_{32}$ as an example, assuming $R_{32}$ is not equal to 0, it can be composed of the following two separable pure state sums (not convex sums):

\begin{align}
   \frac{\left\lvert R_{32}\right\rvert }{2}(R_{32}^1+R_{32}^1)=\begin{pmatrix}
      & 0 & 0 & 0 \\
     0 & 0 & 0 & 0 \\
     0 & R_{32} & 0 & 0 \\
     0 & 0 & 0 & 0
 \end{pmatrix}\\
   R_{32}^1=
      \begin{pmatrix}
           & 1 & 0 & 0 \\
          0 & 0 & 0 & 0 \\
          R_{32}/\left\lvert R_{32}\right\rvert & R_{32}/\left\lvert R_{32}\right\rvert & 0 & 0 \\
          0 & 0 & 0 & 0
      \end{pmatrix}\\
      R_{32}^2=
         \begin{pmatrix}
              & -1 & 0 & 0 \\
             0 & 0 & 0 & 0 \\
             -R_{32}/\left\lvert R_{32}\right\rvert & R_{32}/\left\lvert R_{32}\right\rvert & 0 & 0 \\
             0 & 0 & 0 & 0
      \end{pmatrix}
\end{align}

This means that the vector component corresponding to position $R_{32}$ can be combined using $2\times \frac{\left\lvert R_{32}\right\rvert }{2}$ i.e. the resources of $\left\lvert R_{32}\right\rvert$. Similarly, the next two components at other positions can also be combined using the same method.

In other cases, even if the resources used to construct each component are not $\left\lvert R_{ij}\right\rvert$, we can use a similar method to calculate how many resources it requires, denoted as $N(R_{ij})$.

So in order to construct this target quantum state $\overrightarrow{R}$, we need a total resources $N_R$ of the following quantities of quantum states for combination:

\begin{equation}
   N_R=\sum_{i,j}N(R_{ij}) 
\end{equation}

There are two situations to discuss. Firstly, if $N_R\leqslant 1$, the quantum state is already separable. Secondly, if $N_R>1$, if we shorten the vector corresponding to the quantum state by $N_R$ times, the resources required for the combination will be reduced to 1, that is, the combination will become a convex combination, and the vector will definitely be in the convex hull at this point. In summary, this proves the above conclusion.

Based on the above properties, we can draw a conclusion that if a vector is outside a convex hull, we shorten it by a certain multiple, and it will also enter the convex hull. We will continue to shorten it until the vector is 0, and it will still be in the convex hull.

So we can define a shortening coefficient $\alpha(\alpha\in (0,1])$, whose initial value is 1. If the vector $\overrightarrow{R}$ corresponding to the quantum state is already inside the convex hull, $\alpha$ will not decrease. If it is outside the convex hull, $\alpha$ will eventually decrease to the value corresponding to $\alpha\overrightarrow{R}$ at the boundary of the convex hull. We can relate the multiple that needs to be shortened to how entangled it is and provide our entanglement measurement method. We define the entanglement degree $C(\rho)$ as:

\begin{equation}
C(\rho)=1-\alpha
\end{equation}

In the following sections, I will provide a feasible methods for calculating $\alpha(\overrightarrow{R})$ values.

\section{Querying the value of $\alpha(\overrightarrow{R})$ through the properties of convex hull}

Assuming that for bounded convex geometry in space, we can always create a hyperplane through a point outside the convex geometry to make it tangent to the convex geometry. And according to the properties of convex geometry, all points in convex geometry must be located on one side of the tangent plane\cite{bertsekas2003convex}, because according to the properties of convex geometry, if this is not true, then the plane must not be a tangent hyperplane. And this will provide us with a wealth of information.

\begin{figure}[htbp]
   \centering
   \includegraphics[width=0.3\textwidth]{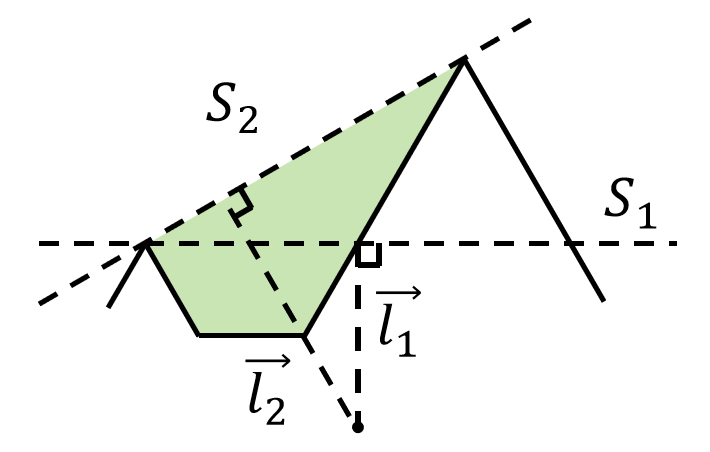}
   \begin{minipage}{0.5\textwidth}
       \centering
       \caption{Super tangent plane of convex hull}
       \label{fig:1}
       \vspace{0.5em} 
       \par \indent S1 cannot be a cross-section of the convex hull surface, as the points in the convex hull appear on both sides of S1 simultaneously. This will result in the dyed part also belonging to the convex hull, making it impossible for the gradients in different directions at the intersection of the tangent plane and the convex hull to be less than or equal to 0 when multiplied by the normal vector of the tangent plane. In contrast, S2 is more like the surface tangent plane of the convex hull.
   \end{minipage}
\end{figure}

For the normal vector $\overrightarrow{l} $, make the hyperplane tangent to the convex hull, and the points $x$ on the tangent plane can be represented as follows:

\begin{equation}
   \overrightarrow{x}  \cdot \overrightarrow{l}=a  
\end{equation}

Assuming that the hyper tangent plane at this point holds $\overrightarrow{x}  \cdot \overrightarrow{l}\leq a$ for all vectors $\overrightarrow{x}$ in the convex hull. Due to the properties of convex geometry, we can conclude that there must be pure states on the cross section, and the value obtained by multiplying the point of the vector corresponding to all pure states on it with the normal vector must be the average.

To demonstrate these two points, we can assume that the tangent plane intersects with convex geometry at only one point, and assume that this point is the vector corresponding to the mixed state. 

\begin{align}
   \overrightarrow{x}=\sum_{i}\lambda_i\overrightarrow{r_i}\\
   \sum_{i}\lambda_i=1, \lambda_i\geqslant 0 
\end{align}

For all pure states used for convex group sums at this time, the maximum value of their dot product must be greater than or equal to the mean. However if it is greater than the average value, it does not meet the requirement that all points are located on one side of the hyper tangent plane, so the maximum value must be equal to the average value. If the maximum value is already equal to the average value, then the minimum value will also be equal to the average value, otherwise the average value will be smaller than itself, which is unreasonable. Therefore, we can conclude that if there is only one point on the tangent plane, it cannot be a mixed state. Because if it is a mixed state vector, then there must be other pure states on the tangent plane, which proves both of these conclusions. Now we know that there must be a pure state on the cross-section.

Firstly, let us assume that the vector corresponding to our target quantum state is $\overrightarrow{R}$, and assume that there is a coefficient $\alpha$ in $(0,1]$. Meanwhile, assuming the normal vector of the tangent plane is $\overrightarrow{l}$, because $\alpha \overrightarrow{R} $ and a pure state $\overrightarrow{r_0}$ are both on this tangent plane, we have $\overrightarrow{l}  \cdot (\alpha \overrightarrow{R}- \overrightarrow{r_0})=0 $, and because of the tangent plane, we have $\overrightarrow{l}\cdot \overrightarrow{r_0}\geqslant  \overrightarrow{l}\cdot\overrightarrow{r}  $. Because $\overrightarrow{l}\cdot \overrightarrow{r_0}$ is the maximum value, and $\overrightarrow{r_0}$ is clearly a correlation function of $\overrightarrow{l}$, we can refer to the result of $\overrightarrow{l}\cdot \overrightarrow{r_0}$ at the point where it is multiplied to its maximum as $F(\overrightarrow{l})$, $F(\overrightarrow{l})=\max_{\{\overrightarrow{r}\}}\overrightarrow{l}\cdot\overrightarrow{r}$. We gradually decrease $\alpha$ until there are no longer $\overrightarrow{l}$ vectors that satisfy these two conditions, and then we can say that the quantum state has reached the boundary of the convex hull.
This measurement method has another purpose, which is to decompose the vector $\overrightarrow{R}$ into a positive weighted sum of pure separable states, and then find a value $\overrightarrow{l}$ that multiplies all decomposed pure states by the same value, and is the maximum value of $\overrightarrow{l}$ multiplied by any pure state. Then, we shorten $\overrightarrow{R}$ to the weighted sum of $\overrightarrow{l}$ to obtain the current $\alpha(\overrightarrow{R})$.

\begin{figure}[htbp]
   \centering
   \includegraphics[width=0.3\textwidth]{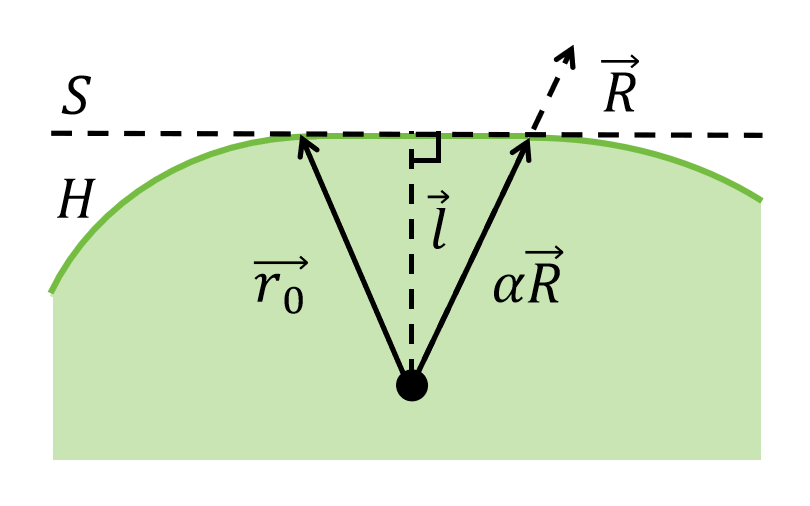}
   \begin{minipage}{0.5\textwidth}
       \centering
       \caption{obtain $\alpha(\overrightarrow{R}) $}
       \label{fig:2}
       \vspace{0.5em} 
       \par \indent Now $H$ represents the boundary of the convex hull, $S$ represents the surface section of the convex hull at this time, gradually reducing $\alpha$ to reach the boundary of the convex hull.
   \end{minipage}
\end{figure}

This can be achieved through a program. Firstly, we traverse the vector $\overrightarrow{l}$ and find the maximum value that corresponds to each $\overrightarrow{l}$ multiplied by $\overrightarrow{l}$. Then, we calculate the value of $\overrightarrow{l} \cdot \overrightarrow{R}$ and compare the two to obtain the current value, which we call $\beta $. If $\beta<0 $ or $\beta\geqslant 1$, we do not change the value of $\alpha $ and continue to traverse $\overrightarrow{l}$. If it is between 0 and 1, we then Compare it with $\alpha $, where the initial value of $\alpha $ is 1. If $\beta<\alpha $, we replace $\alpha $ with $\beta $. If it is greater than or equal to $\overrightarrow{l}$, we continue to iterate. In the end, we will obtain the minimum value of $\alpha$.

\begin{equation}
   \alpha(\overrightarrow{R})=\min_{\{\overrightarrow{l}\}}\frac{F(\overrightarrow{l})}{\overrightarrow{l}\cdot \overrightarrow{R}},\alpha\in (0,1]
\end{equation}

Meanwhile, based on the equation (3), we can conclude that this equation is equal to:

\begin{equation}
   \alpha(\rho)=\min_{\{\varrho\}}\frac{F(\varrho)-\frac{1}{d}}{Tr(\varrho\cdot \rho)-\frac{1}{d}},\alpha\in (0,1]
\end{equation}

Where $\varrho$ is the density matrix of any quantum state in that dimension, and similarly, we define $F(\varrho)$ as the maximum value within and after quantum state $\varrho$ and pure separable state $|\psi^{S} \rangle$, $F(\varrho)=\max_{\{|\psi^{S} \rangle\}}\langle\psi^{S}|\varrho|\psi^{S} \rangle $, it is worth mentioning that there are indeed relevant studies on this issue\cite{{biswas2023fidelity}}. So at this point, assuming that the final quantum state $\varrho$ is $\varrho_0$, our entanglement degree $C(\rho)$ is equal to:

\begin{equation}
   C(\rho)=1-\alpha(\rho)=\frac{Tr(\varrho_0\cdot \rho)-F(\varrho_0)}{Tr(\varrho_0\cdot \rho)-\frac{1}{d}},C\in [0,1)
\end{equation}

The maximum value of entanglement in this method is not 1. In practical use, the overall result can be divided by the maximum value in this case to achieve a maximum value of 1.

This measurement method has another use, which is to decompose the vector $\overrightarrow{R}$ into a positive weighted sum of pure separable states $\overrightarrow{r_i}$:

\begin{equation}
   \overrightarrow{R}=\sum_{i}b_i \overrightarrow{r_i}, b_i\geqslant 0
\end{equation}

And then find a vector $\overrightarrow{l}$ that multiplies all decomposed pure states $\overrightarrow{r_i}$ to the same value, and is the maximum value of $\overrightarrow{l}$ multiplied by any pure state $\overrightarrow{r}$. This indicates that at this point, $\overrightarrow{l}$ is the normal vector of all planes of the convex hull, and $\overrightarrow{r_i}$ is the pure separable state on the tangent plane. This indicates that if the vector of the quantum state is shortened by $\sum_{i}b_i$ times, it will be on the same convex tangent plane as the pure state separable state $\overrightarrow{r_i}$. Then, we shorten $\overrightarrow{R}$ to $\alpha(\overrightarrow{R})$ at this point, $\alpha$ now is:

\begin{equation}
   \alpha=\frac{1}{\sum_{i}b_i}
\end{equation}

\section{Example of Results}

Here is an example of a Wenner state\cite{{werner1989quantum}} for a 2-qubit GHZ state\cite{GHZ89}:

\begin{equation}
   | \psi_{GHZ} \rangle=cos\theta| 00 \rangle+sin\theta| 11\rangle
\end{equation}

\begin{equation}
   R(| \psi_{GHZ} \rangle )=\begin{pmatrix}
         & 0 & 0 & cos2\theta \\
       0 & sin2\theta & 0 & 0 \\
       0 & 0 & -sin2\theta & 0 \\
       cos2\theta & 0  & 0 & 1 
   \end{pmatrix}
 \end{equation}

 Discuss the situation where $sin2\theta\geqslant 0$, we can directly provide the normal vector $\overrightarrow{l}$ of the section at this time:

 \begin{equation}
   l=\begin{pmatrix}
         & 0 &  0 & 0 \\
       0 & 1 &  0 & 0 \\
       0 & 0 & -1 & 0 \\
       0 & 0 &  0 & 1 
   \end{pmatrix}
 \end{equation}

 Based on the 2qubit pure state vector obtained in the previous text, we can obtain the form obtained by multiplying $\overrightarrow{l}$ and $\overrightarrow{r}$ points as follows:

\begin{equation}
   \overrightarrow{l}\cdot\overrightarrow{r}=v_x u_x-v_y u_y+v_z u_z\leqslant 1
\end{equation}

We can now split $\overrightarrow{R}$ into 6 vectors multiplied by 1 with point $\overrightarrow{l}$:

\begin{equation}
   R=\frac{sin2\theta}{2}(R_1+R_2+R_3+R_4)+cos^2\theta R_5+sin^2\theta R_6
\end{equation}

\begin{align}
   R_1=\begin{pmatrix}
         & 1 &  0 & 0 \\
       1 & 1 &  0 & 0 \\
       0 & 0 &  0 & 0 \\
       0 & 0 &  0 & 0 
   \end{pmatrix}\\
   R_2=\begin{pmatrix}
       & -1 &  0 & 0 \\
    -1 &  1 &  0 & 0 \\
     0 &  0 &  0 & 0 \\
     0 &  0 &  0 & 0 
   \end{pmatrix}\\
   R_3=\begin{pmatrix}
         & 0 & -1 & 0 \\
       0 & 0 &  0 & 0 \\
       1 & 0 & -1 & 0 \\
       0 & 0 &  0 & 0 
   \end{pmatrix}\\
   R_4=\begin{pmatrix}
      & 0 &  1 & 0 \\
    0 & 0 &  0 & 0 \\
   -1 & 0 & -1 & 0 \\
    0 & 0 &  0 & 0 
   \end{pmatrix}\\
   R_5=\begin{pmatrix}
      & 0 &  0 & 1 \\
    0 & 0 &  0 & 0 \\
    0 & 0 &  0 & 0 \\
    1 & 0 &  0 & 1 
   \end{pmatrix}\\
   R_6=\begin{pmatrix}
      & 0 &  0 & -1 \\
    0 & 0 &  0 &  0 \\
    0 & 0 &  0 &  0 \\
   -1 & 0 &  0 &  1 
   \end{pmatrix}
\end{align}

So we can calculate the $\alpha(\overrightarrow{R})$ value at this time as:

\begin{equation}
   \alpha(\overrightarrow{R})=\frac{1}{1+2sin2\theta}
\end{equation}

The Wenner state for a 2-qubit GHZ state can be written as\cite{werner1989quantum}:

\begin{equation}
   \rho_W=V| \psi_{GHZ} \rangle\langle\psi_{GHZ}|+(1-V)\frac{1}{4}I,V\in [0,1]
\end{equation}

Based on our conclusion, we can clearly know that max $V$ at this point is the $\alpha(\overrightarrow{R})$ we calculated.

It can be seen that the boundary value obtained by the PPT method is consistent:

\begin{align}
   \frac{1}{2}sin2\theta V\leqslant \frac{1}{4}(1-V)\\
   V\leqslant \frac{1}{1+2sin2\theta}
\end{align}

This indicates that when the value of $V$ is the value of $\alpha$ we calculated, it happens to be at the boundary between entanglement and non entanglement, which is consistent with the idea of convex hull boundary in our work.

\section{Conclusion and Prospect}

In this work, we introduce another approach to entanglement measurement, starting from the point that the set of separable states is a convex hull of separable pure states, and analyze the entanglement problem based on the properties of the convex hull. At the same time, we provide a brief proof of the feasibility of this entanglement measurement method. Afterwards, we provided two calculation schemes. In cases where we cannot determine the corresponding l, we should use numerical calculations. However, when we already know the corresponding $\overrightarrow{l}$  and $\overrightarrow{r}$, we can use the second non numerical analytical scheme. 

Although we have only provided examples of the 2-qubit scenario in the text, this method is not only applicable to the 2-qubit form, but also satisfies various types of quantum entanglement metrics. Because the properties of convex hull do not change with entanglement or separability. In the future, we will also use the first numerical method to provide more data support for our entanglement measurement method.


\begin{thebibliography}{0}%
\makeatletter
\providecommand \@ifxundefined [1]{%
 \@ifx{#1\undefined}
}%
\providecommand \@ifnum [1]{%
 \ifnum #1\expandafter \@firstoftwo
 \else \expandafter \@secondoftwo
 \fi
}%
\providecommand \@ifx [1]{%
 \ifx #1\expandafter \@firstoftwo
 \else \expandafter \@secondoftwo
 \fi
}%
\providecommand \natexlab [1]{#1}%
\providecommand \enquote  [1]{``#1''}%
\providecommand \bibnamefont  [1]{#1}%
\providecommand \bibfnamefont [1]{#1}%
\providecommand \citenamefont [1]{#1}%
\providecommand \href@noop [0]{\@secondoftwo}%
\providecommand \href [0]{\begingroup \@sanitize@url \@href}%
\providecommand \@href[1]{\@@startlink{#1}\@@href}%
\providecommand \@@href[1]{\endgroup#1\@@endlink}%
\providecommand \@sanitize@url [0]{\catcode `\\12\catcode `\$12\catcode `\&12\catcode `\#12\catcode `\^12\catcode `\_12\catcode `\%12\relax}%
\providecommand \@@startlink[1]{}%
\providecommand \@@endlink[0]{}%
\providecommand \url  [0]{\begingroup\@sanitize@url \@url }%
\providecommand \@url [1]{\endgroup\@href {#1}{\urlprefix }}%
\providecommand \urlprefix  [0]{URL }%
\providecommand \Eprint [0]{\href }%
\providecommand \doibase [0]{http://dx.doi.org/}%
\providecommand \selectlanguage [0]{\@gobble}%
\providecommand \bibinfo  [0]{\@secondoftwo}%
\providecommand \bibfield  [0]{\@secondoftwo}%
\providecommand \translation [1]{[#1]}%
\providecommand \BibitemOpen [0]{}%
\providecommand \bibitemStop [0]{}%
\providecommand \bibitemNoStop [0]{.\EOS\space}%
\providecommand \EOS [0]{\spacefactor3000\relax}%
\providecommand \BibitemShut  [1]{\csname bibitem#1\endcsname}%
\let\auto@bib@innerbib\@empty
\end{thebibliography}%


\begin{thebibliography}{99}


   \bibitem{EPR1935}
   A. Einstein, B. Podolsky, and N. Rosen,
   \href{https://doi.org/10.1103/PhysRev.47.777}{Phys. Rev. \textbf{47}, 777 (1935)}.

   \bibitem{bell1964}
   J. S. Bell,
   \href{https://doi.org/10.1103/PhysicsPhysiqueFizika.1.195}{Physics  \textbf{1}, 195 (1964)}.

   \bibitem{aspect1982experimental}
   A. Aspect, J. Dalibard, and G. Roger{\'e}rard,
   \href{https://journals.aps.org/prl/abstract/10.1103/PhysRevLett.49.1804}{Phys. Rev. Lett  \textbf{49}, 1804 (1982)}.

   \bibitem{vedral1997quantifying}
   V. Vedral, M. B. Plenio, M. A. Rippin, and P. L. Knight,
   \href{https://journals.aps.org/prl/abstract/10.1103/PhysRevLett.78.2275}{Phys. Rev. Lett  \textbf{78}, 2275 (1997)}.

   \bibitem{wootters2001entanglement}
   W. K. Wootters,
   \href{https://www.rintonpress.com/journals/qic-1-1/eof2.pdf}{Quantum Inf. Comput.  \textbf{1}, 27--44 (2001)}.

   \bibitem{peres1996separability}
   A. Peres,
   \href{https://journals.aps.org/prl/abstract/10.1103/PhysRevLett.77.1413}{Phys. Rev. Lett  \textbf{77}, 1413 (1996)}.

   \bibitem{horodecki1996necessary}
   M. Horodecki, P. Horodecki and R. Horodecki,
   \href{https://citeseerx.ist.psu.edu/document?repid=rep1&type=pdf&doi=f829c395a281fde3395edaff91469190aa77dbc3}{Phys. Rev. A  \textbf{223},(1996)}.

   \bibitem{perlmutter2015positivity}
   E. Perlmutter, M. Rangamani, and M. Rota, 
   \href{https://arxiv.org/abs/1506.01679}{arXiv preprint arXiv:1506.01679,(2015)}.

   \bibitem{banerji2014entanglement}
   A. Banerji, R. P. Singh and A. Bandyopadhyay,
   \href{https://www.sciencedirect.com/science/article/abs/pii/S0030401814004908}{Optics Communications \textbf{330}, 85--90 (2014)}.

   \bibitem{li2024quantum}
   J. X. Li, H. M. Yao, S. M. Fei, and Z. B. Fan and H. T. Ma,
   \href{https://journals.aps.org/pra/abstract/10.1103/PhysRevA.109.052426}{Phys. Rev. A  \textbf{109}, 052426 (2024)}.

   \bibitem{diez2021quantum}
   P. D{\'\i}ez-Valle, D. Porras and J. J. Garc{\'\i}a-Ripoll,
   \href{https://journals.aps.org/pra/abstract/10.1103/PhysRevA.104.062426}{Phys. Rev. A  \textbf{104}, 062426 (2021)}.

   \bibitem{lin2023quantifying}
   X. D. Lin, Z. Y. Chen and Z. H. Wei,
   \href{https://journals.aps.org/pra/abstract/10.1103/PhysRevA.107.062409}{Phys. Rev. A  \textbf{107}, 062409 (2023)}.

   \bibitem{greenwood2023machine}
   A. C. B. Greenwood, L. T. H. Wu, E. Y. Zhu, B. T. Kirby and L. Qian, 
   \href{https://journals.aps.org/prapplied/abstract/10.1103/PhysRevApplied.19.034058}{Phys. Rev. A  \textbf{19}, 034058 (2023)}.

   \bibitem{chen2002degree}
   J. L. Chen, L. B. Fu, A. A. Ungar and X. G. Zhao,
   \href{https://journals.aps.org/pra/abstract/10.1103/PhysRevA.64.050101}{Phys. Rev. A  \textbf{65}, 044303 (2002)}.

   \bibitem{bertsekas2003convex}
   D. Bertsekas, A. Nedic and A. Ozdaglar,
   \href{https://journals.aps.org/pra/abstract/10.1103/PhysRevA.64.050101}{Phys. Rev. A  \textbf{1}, (2003)}.

   \bibitem{pauli1988quantenmechanik}
   W. Pauli,
   \href{https://link.springer.com/chapter/10.1007/978-3-322-90270-2_32}{(1988)}.

   \bibitem{bloch1946nuclear}
   F. Bloch,
   \href{https://journals.aps.org/pr/abstract/10.1103/PhysRev.70.460}{Phys. Rev  \textbf{70}, 460 (1946)}.

   \bibitem{biswas2023fidelity}
   G. Biswas, S. H. Hu, J. Y. Wu, D. Biswas and A.Biswas,
   \href{https://iopscience.iop.org/article/10.1088/1402-4896/ad4f2f/meta}{Physica Scripta, (2023)}.

   \bibitem{werner1989quantum}
   R. F. Werner,
   \href{https://journals.aps.org/pra/abstract/10.1103/PhysRevA.40.4277}{Phys. Rev. A  \textbf{40}, 4277 (1989)}.

   \bibitem {GHZ89}
   D. M. Greenberger, M. A. Horne, and A. Zeilinger,
   in \textit{Bell's Theorem, Quantum Theory, and Conceptions of the Universe}, edited by M. Kafatos (Kluwer, Dordrecht, 1989), p. 69.

    


\end{thebibliography}
\end{document}